\documentclass[12pt, preprint]{aastex}
\usepackage{graphicx}

\shorttitle{An imaging study of a complex solar coronal radio eruption}

\shortauthors{Feng et al.}

\begin{document}

\title{An imaging study of a complex solar coronal radio eruption}

\author{S. W. Feng\altaffilmark{1,2}, Y. Chen\altaffilmark{1}, H. Q. Song
\altaffilmark{1}, B. Wang\altaffilmark{1}, X. L. Kong \altaffilmark{1} }
\altaffiltext{1}{Shandong Provincial Key Laboratory of
Optical Astronomy and Solar-Terrestrial Environment, and Institute of Space Sciences,
Shandong University, Weihai, Shandong 264209, China; yaochen@sdu.edu.cn}

\altaffiltext{2}{State Key Laboratory of Space Weather, Chinese Academy of Sciences, Beijing 100190, China}

\begin{abstract}

Solar coronal radio bursts are enhanced radio emission excited by
energetic electrons accelerated during solar eruptions, studies on
which are important for investigating the origin and physical
mechanism of energetic particles and further diagnosing coronal
parameters. Earlier studies suffered from a lack of simultaneous
high-quality imaging data of the radio burst and the eruptive
structure in the inner corona. Here we present a study on a
complex solar radio eruption consisting of a type II and three
reversely-drifting type III bursts, using simultaneous EUV and
radio imaging data. It is found that the type II burst is closely
associated with a propagating and evolving CME-driven EUV shock
structure, originated initially at the northern shock flank and
later transferred to the top part of the shock. This source
transfer is co-incident with the presence of shock decay and
enhancing signatures observed at the corresponding side of the EUV
front. The electron energy accelerated by the shock at the flank
is estimated to be $\sim$ 0.3 c by examining the imaging data of
the fast-drifting herringbone structure of the type II burst. The
reversely-drifting type III sources are found to be within the
ejecta and correlated with a likely reconnection event therein.
Implications on further observational studies and relevant
space-weather forecasting techniques are discussed.
\end{abstract}


\section{Introduction}
Coronal radio bursts at metric wavelength are observational
manifestation of energetic electrons that are accelerated in the
solar atmosphere. These radio bursts, carrying valuable
information about how energetic particles are accelerated and the
underlying emission mechanism, can be used to infer parameters of
coronal magnetic field and plasmas, which otherwise remain
difficult to measure with other techniques.

Among various types of coronal radio bursts, the type II, narrow
and slowly-drifting bands on the solar radio dynamic spectrum, is
believed to be excited by energetic electrons accelerated at
coronal shocks. It has been discovered (Allen 1947; Payne-Scott
1947) and studied for nearly 70 years, yet major issues, such as
where and how the electrons are accelerated by a shock and how
some fine structures (e.g., splitting bands, see latest studies by
Vasanth et al. 2014 and  Du et al. 2014, 2015, and herringbones,
see e.g., Cairns \& Robinson 1987) are formed, remain unresolved.
Regarding the relative source location on the shock surface,
earlier studies claimed that both the nose front and the flank of
a shock can be sources of type IIs (Reiner et al. 2003; Cho et al.
2007, 2008; Feng et al. 2012, 2013; Kong et al. 2012; Chen et al.
2014). These studies, providing valuable insights into the type II
origin, are mostly based on non-imaging spectrometric data of the
radio burst and low-cadence white-light data of the coronal mass
ejection (CME). Considering metric radio bursts occur within 2-3
solar radii from the solar center (as inferred from their
frequencies) with a relatively short duration of several to $\sim$
10 minutes, direct imaging data of the metric t-II source are not
available for most events analyzed earlier.

This situation has changed with the availability of the
high-cadence high-sensitivity Atmospheric Imaging Assembly
on board the Solar Dynamics Observatory (AIA/SDO) instrument
(Lemen et al. 2012; Pesnell et al. 2012) with EUV passbands
covering a broad range of plasma temperatures, together with the
ground-based solar radio heliograph instrument working in the
metric wavelength, especially the Nan\c{c}ay Radio Heliograph
(NRH, Kerdraon \& Delouis 1997). In latest studies, these two sets
of imaging data have been combined to reveal more insights into
the physics of metric radio bursts, not only type II (Feng et al.
2015; Zimovets, et al. 2015; Eselevich, et al. 2015), but also
type IV radio bursts (Tun \& Vourlidas 2013, Bain et al. 2014,
Vasanth et al. 2016) among others. For example, Feng et al. (2015)
investigated the source regions of a three-lane type II burst, and
found that all lanes are located at the shock flank with two lanes
on the southern part and the other one from a distinct location on
the northern part, consistent with the result obtained by Zimovets
et al. (2015) on the same event.

Latest numerical studies on the role of the large-scale closed
magnetic field in shock-electron acceleration suggest that the
relative curvature of the shock and the closed field across which
the shock is crossing is important to the efficiency of electron
acceleration (Kong et al. 2015, 2016). They further deduced that
at low shock altitudes, when the shock is more curved than the
interacting closed magnetic field, the electrons are mainly
accelerated at the shock flank; at higher altitudes, when the
shock is less curved, the electrons are mainly accelerated at the
shock nose around the top of the closed field lines. Both
the local shock geometry (being quasi-parallel or
quasi-perpendicular, with the latter being expected to be more
efficient in electron acceleration (Wu et al. 1984)) and the
trapping effect of closed field lines play a role in the described
process. This predicts a shift of efficient electron acceleration
site along the shock front and a transfer of corresponding radio
emitting sources from the flank to the top during the shock
propagation. Such a transfer of radio sources along a coronal
shock has not been reported, to the best of our knowledge.

Another topic of particular interest here is related to the type
III  radio burst, which corresponds to fast drifting features on
the dynamic spectrum, being one of the \textbf{best} understood
type of coronal radio bursts. The type III burst is thought to be
excited by fast electron beams propagating along open or open-like
(i.e., large-scale closed) field lines via the classical plasma
emission mechanism (Ginzburg \& Zhelezniakov 1958, Melrose 1980).
The energetic electrons are usually believed to be accelerated by
magnetic reconnection accounting for solar flares at the base of
the corona. Besides the usual type III burst drifting from higher
to lower frequencies, reversely-drifting or even bi-directional
type III bursts are sometimes observed, with the other component
that drifts from lower to higher frequencies (Aschwanden et al.
1993, 1995; Robinson \& Benz 2000; Ning et al. 2000; Tan et al.
2016). These type IIIs are regarded as a sensitive tool to
diagnose the physical conditions around the magnetic energy
release site where reconnection and particle acceleration take
place, \textbf{studies on which suffer from the general lack of
simultaneous radio and EUV imaging data.}

Here we present an observational study on a complex coronal radio
burst characterized by both type II and reversely-drifting type
III bursts. It is associated with a solar eruption observed from
the limb in the AIA field of view (FOV), and the accompanying EUV
shock structure is clearly observable. Simultaneous NRH data at
several metric wavelengths from 445 to 228 MHz are also available.
Interesting results about the radio sources and its connection to
the underlying eruptive processes are obtained.

\section{Observational data and event overview}
For the dynamic spectra we used the data recorded by the ORFEES
spectrometer at the Nan\c{c}ay observatory at higher frequency (1
GHz - 140 MHz) and the San Vito data of the Radio Solar Telescope
Network (RSTN) at lower frequency (140 - 25 MHz). The associated
eruption is imaged by the AIA/SDO in different EUV passbands with
a 12s cadence and a pixel size of 0.6'', and the radio sources are
imaged by the NRH from 150-445 MHz with spatial resolution
depending on the imaging frequency ($\sim$ 1'-2' at 445 MHz,
decreasing to 5'-6' at 150 MHz).

The event is associated with an M5.9 class flare on 24 August
2014, which starts at 12:00 UT, peaks at 12:17 UT and ends at
12:25 UT according to the GOES SXR data. It is from the NOAA
active region (AR 12151, S09E76) and accompanied by a filament
eruption on the limb which further evolves into a CME with a
linear speed of 550 km s$^{-1}$ according to the Large Angle
Spectroscopic Coronagraph (Brueckner et al. 1995) C2 data. From
AIA, the erupting filament presents a nice $\Omega$-shape
morphology, indicating the occurrence of the flux rope kink
instability (T\"{o}r\"{o}k \& Kliem 2005). Song et al. (2015)
presented a dynamical analysis on the event and found that the
filament experiences a two-step fast acceleration phases with the
first one not and the second one corresponding to a significant
increase of HXR profiles. They thus suggested that the first
acceleration phase is driven by the energy release dominated by
ideal flux rope instability and the second phase dominated by
flare-related magnetic reconnection. The amounts of energy release
of the two steps are comparable, consistent with earlier numerical
simulation (Chen et al. 2007).

Here we focus on the AIA EUV wave fronts which are driven by and
propagate ahead of the eruptive filament. See Figure 1 for the
evolutionary sequence of the wave structure observed at 211 {\AA}.
The front has a circular shape in the projection plane. Its
temporal evolution can be viewed from Figure 1, the accompanying
online animation, and the distance-time maps along the two slices
S1 and S2 (see Figure 2). Before 12:11:16 UT (see Figure 1a-1b),
the front along S1 exhibits a complex multi-\textbf{shell}
structure, while at $\sim$ 12:12 UT and later (Figure 1c-1h), the
multi-\textbf{shell} structure is pushed forward by the ejecta and
steepens into a single-\textbf{shell} bright structure. This may
indicate the formation of the shock there, as confirmed later by
the presence of co-spatial type II radio sources. While examining
the EUV wave evolution at the lateral side, we find that at $\sim$
12:10 UT, the wave there is already single-\textbf{shell}, a
result of earlier super-radial lateral expansion of the eruptive
structure. This may indicate the formation of the shock structure
is about two minutes earlier at the flank than at the top part.
After 12:12:23 UT (Figure 1d), the EUV wave at the flank near the
Sun ($<$ 1.1 R$_\odot$) becomes diffuse and not recognizable
later, indicating possible decay of the shock there. The EUV wave
morphological evolution is in line with the radio data to be
presented. The propagation speeds as measured from the slice maps
are $\sim$ 120 km s$^{-1}$ for the EUV wave flank from 12:10 to
12:12 UT, slower than the speed at the nose front of the EUV wave
($\sim$ 600 km s$^{-1}$) as measured from 12:12 to 12:14 UT.

The dynamic spectra of the event is shown in Figure 3a. It is a
complex event consisting of several types of radio bursts. From
high to low frequencies, there are the first episode of the type
III ($\sim$ 12:03 UT) just 3 minutes after the flare start, its
second episode (from 12:10 - 12:16 UT) that embeds within the
type-IV continuum. At their low-frequency end, the type III and IV
bursts are enveloped by a narrow and relatively slowly-drifting
band of type II. At the lower end of the type II, a cluster of
type I with sporadic brightenings appearing from 200 to 250 MHz
after 12:10 UT. At lower frequencies, the emission is dominated by
a second episode of type II with split bands and
fundamental-harmonic branches.

\section{Imaging the type II radio burst and its herringbone structure}

Now we focus on the bursts shown in Figure 3b, corresponding to
the box region of Figure 3a. We can see that the high-frequency
type II band is characterized by discontinuous and fragmented
emission, consisting of many fast drifting herringbone structures.
The NRH total brightness temperatures ($T_{B}$) at frequencies
from 445 to 228 MHz, normalized by the corresponding maximum
$T_{B}$, are superposed to show the nice correlation between the
NRH data and the spectral data. Along the type II band, we select
eight spectral points (indicated by crosses in Figure 3b) at which
the NRH $T_{B}$ peaks. The corresponding type II sources have been
superposed onto the AIA data in Figure 1.

The most important radio-source information \textbf{visible in}
Figure 1 is the shift of type II sources with time. From the first
three panels and before 12:11:59 UT, the type II sources are
located at the flank of the EUV wave, indicating the shock
presence there. In Figure 1(d), the type II source appears at the
top part of the EUV wave and stays there later. The time of the
source shift occurs between 12:11:50 to 12:12:20 UT. During this
interval, the type II spectrum presents some discontinuity with a
sharp change of frequency, possibly associated with the shift. As
mentioned earlier, the source shift is in line with the EUV wave
morphological evolution, as shown in Figures 1 and 2. There we see
that before 12:12 UT the flank of the EUV wave (corresponding to
the S2 map), where the type II sources were located then, shows a
strong shock-like single-\textbf{shell} bright structure, while at
the top part (corresponding to the S1 map) the
multi-\textbf{shell} structure steepens into a
single-\textbf{shell} shock-like feature after 12:12 UT,
co-incident with the radio source jump from the flank to top.

As mentioned the type II backbone consists of fast drifting
herringbone structures, which are suggested to be excited by fast
energetic electrons accelerated at and escaped from the shock
(Cairns \& Robinson 1987). With the source images of the
herringbone at different frequencies, one can estimate the speed
of shock-released electrons. The herringbone structure usually
occupies only a limited frequency range, so it is in general
difficult to do this. Nevertheless, here we identify one such
herringbone structure that is imaged by NRH at two frequencies at
408 and 327 MHz (see the red box region in Figure 3). The
corresponding spectral data and NRH source contours are shown in
Figure 4. Assuming that the peaks of the radio flux
correspond to the same beam of electrons, we can track their
motion and estimate their speed. At 12:12:20 UT the electrons (and
the herringbone source at 408 MHz) are at the flank, while $\sim$
1.5s later (inferred from the timing difference of the two plus
signs) the electrons move to the top part of the ejecta, so does
the herringbone source at 327 MHz. It is not possible to determine
the exact trajectory of the electron beam, yet we can obtain a
lower estimate of the electron speed by assuming the distance
between the source centroids ($\sim$ 200 arcsecs) to be the moving
distance of the beam within the 1.5s interval. This gives a speed
of $\sim$ 0.3 c. Note that this should be regarded as a rough
estimate mainly due to the limitation of the NRH spatial
resolution. The herringbone structure appears during the source
transfer (or jump) process and the AIA-observed shock decay at the
flank, implicating a possible link among these phenomena.

\section{Imaging the reversely-drifting type IIIs and the associated magnetic reconnection}

Now we focus on the type III bursts within the white (blue) box
region of Figure 3. We re-draw the spectrum in Figure 5a and 5b,
and superpose the NRH sources at 432 MHz onto the AIA 304 {\AA}
images (see Figure 5h). We see that there are three sets of type
III, each imaged by the NRH at channels of 408, 432, and 445 MHz.
Examining the fluxes, we see that all these type IIIs present
likely reversely-drifting characteristics, and for the middle one
there seems to exist bi-directional drifting signature with the
frequency at which the bi-directional drift starts being around
$\sim$ 450-500 MHz.

From Figure 5h, we see that the 432 MHz sources of these type IIIs
are located within the ejecta, which are spatially separated from
the overlying EUV wave. This implies the emitting electrons may be
accelerated by processes taking place within the ejecta and
unrelated to the shock. To verify this \textbf{assumption}, we
examine the AIA data at various wavelengths and indeed find a
possible reconnection process there.

The blue arrows in Figure 5c and 5f point at a pair of parallel
and straight filamentary structures that are stretched outward by
the eruption. In Figure 5d and 5g, a retreating bright knot
appears and the two straight structures get kinked (see the
arrows). Later in Figure 5e and 5h, the left filamentary structure
seems to get connected to a different part of the ejecta. These
sudden dynamical morphology change, most clearly viewed from the
accompanying movie, indicates the occurrence of reconnection,
which is co-incident with the appearance of the three
reversely-drifting type IIIs. These observations support that the
type IIIs are excited by energetic electrons accelerated by
reconnections taking place within the ejecta. To estimate
the speed of type-III emitting electrons, we assume the
reconnection site is reasonably close to the source centroids of
the 432 MHz emission, as supported by our deduction of the
frequency range at which the bi-directional type III burst starts.
Then, the height of the reconnection site is estimated to be
$\sim$ 100 arcsec, taken from the mean height of the three source
centroids (see Figure 5h). The electron propagation time from the
reconnection site to the solar limb can be estimated by measuring
the duration of the three sets of reversely-drifting type III,
this gives us an interval of 1-2 s (see the three pairs of short
vertical arrows plotted on top of Figure 5a-5b). Thus, we get the
electron speed to be $\sim$ 0.1 - 0.2 c. This is at the energy
level comparable to those accounting for the type II herringbone.

\section{Summary and discussion}

We analyzed a complex solar radio burst with a combining analysis
on the AIA-EUV images and the NRH radio sources at multiple
metric-wavelengths. We focused on the high-frequency type II and
III bursts. The type II burst is characterized by discontinuous
emissions and fast drifting herringbone structures. We find that
the first part (before 12:12 UT) of the type II burst is
originated from the lateral side of the EUV wave and the later
part is originated from its top, indicating a transfer of radio
sources. This transfer, taking place within $\sim$ 30 s, is
consistent with AIA signatures of the EUV wave steepening
(indicative of shock generation) and diffuse (indicative of shock
\textbf{decay}) at its lateral side and its top. Measurements on
one embedding herringbone source allow us to estimate the speed
and energy of electrons released from the shock, $\sim$ 0.3 c
($\sim$ 25 keV). For the type III bursts, we provide observational
evidence to support that they correspond to sunward-drifting
electron beams accelerated via magnetic reconnection occurring
within the ejecta (one of them also presents likely signature of
anti-sunward drifting, i.e., being a bi-directional one). The
speed of these electron beams are estimated to be 0.1 - 0.2 c.

The observation of the type II source transfer along a single type
II drifting band has a significant implication to relevant studies
using the type II spectral drift to infer coronal shock
properties, especially the shock propagation speed. These studies
assume the type II sources are carried outwards by the shock along
a certain direction, and infer the shock speed with a prescribed
coronal density model. On this basis, several space weather
forecasting schemes are constructed (Smith \& Dryer 1990, Fry et
al. 2001, Zhao \& Feng 2015). Here we show that the type II
sources along a single band can move from the flank side to the
top front, thus a single density model is not applicable to this
kind of event. Note that limits of using type II spectral data
with a fixed density model have been pointed out by many earlier
studies (e.g., Gopalswamy et al. 2009, Liu et al. 2009, Cho et al.
2011), and several studies demonstrated how the type II spectral
shape can be affected by the coronal density structures along the
shock-radio source path, resulting in special spectral
characteristics such as spectral bumps (Feng et al. 2012, 2013) or
break (Kong et al. 2012).

As mentioned in the introduction, Kong et al. (2016) predicted
that the source of efficient electron acceleration and possible
type II burst can shift from the shock flank to its top part,
mainly due to the change of the relative curvature of the
spherical shock and the closed magnetic field lines. In that
study, the dynamical evolution of the shock structure and its
coupling with the coronal background are not considered
self-consistently. Here from the data we show that the shock
dynamical evolutionary process such as its enhancement (or
generation) and decay also plays a critical role in determining
the electron acceleration and the type II source motion.
Besides the above factor, the change of the shock geometry
may also play a role here. Initially, the surrounding field lines
near the sun at the northern flank may be radial and thus parallel
to the shock front (i.e., being a quasi-perpendicular shock,
expected to be efficient in electron acceleration), while later
the shock top may become quasi perpendicular if the shock then is
sweeping across some overlying closed magnetic loops. This likely
change of geometry at different side of the shock during its
evolution is also in line with the observed type-II source
transfer.

The type III bursts of this event are enveloped by the type II
bands, with a close temporal-spectral coincidence. This makes one
suspect that these bursts are physically connected with a common
origin at the EUV shock. However, we show that the type III burst
is very likely related to the reconnection process occurring
within the ejecta during the eruption. The reconnection in the
corona results in bi-directional electron beams, with the
outward-flowing electrons accounting for the NRH sources around
400 MHz, and the sunward-flowing electrons accounting for the
high-frequency part. The latter may further excite X-ray emission
when colliding with the lower solar atmosphere. To verify this, we
examine the Ramaty High Energy Solar Spectroscopic Imager (RHESSI;
Lin et al. 2002) data and found that the type IIIs are co-incident
with strong rising X-ray emissions with oscillating light curves,
it is thus not possible to identify the enhanced X-ray emission
possibly associated with the type III emitting electrons. Further
studies demand imaging data at higher-than-NRH frequencies for a
better identification of the acceleration site and trajectory of
electron beams.

\acknowledgements We gratefully acknowledge the usage of data from
the SDO-SOHO-RHESSI spacecraft and from the Nan\c{c}ay observatory
and the RSTN. S. Feng thank Dr. Hui Fu for helpful discussion.
This work was supported by grants NSBRSF 2012CB825601, NNSFC-CAS
U1431103, and NNSFC 41331068, 41274175, 11503014, NSF of Shandong
Province (ZR2013DQ004, ZR2014DQ001), and Project Supported by the
Specialized Research Fund for the State Key Laboratories.

\newpage
\begin{figure}
 \includegraphics[width=1.0\textwidth]{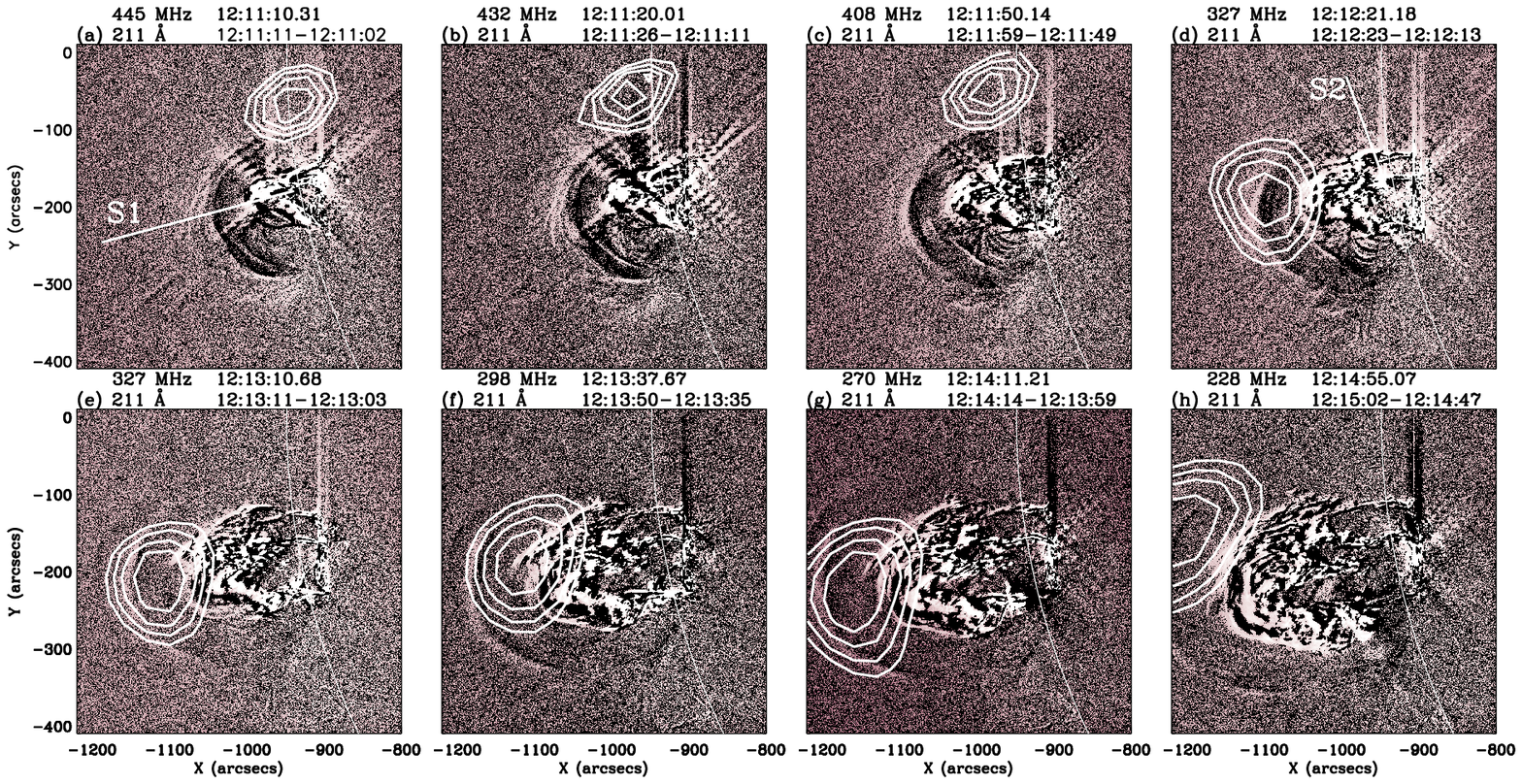}
\caption{Temporal evolution of the EUV wave fronts, overlaid by
the NRH type-II sources. The two slices S1 and S2 are for the
distance-time maps to be shown in Figure 2. The type II sources
are presented by contours levels of 95\%, 90\%, 85\%, and 80\% of
the maximum $T_{B}$ of each NRH image. An accompanying movie is
available online.} \label{Fig:fig1}
\end{figure}

\begin{figure}
 \includegraphics[width=1.0\textwidth]{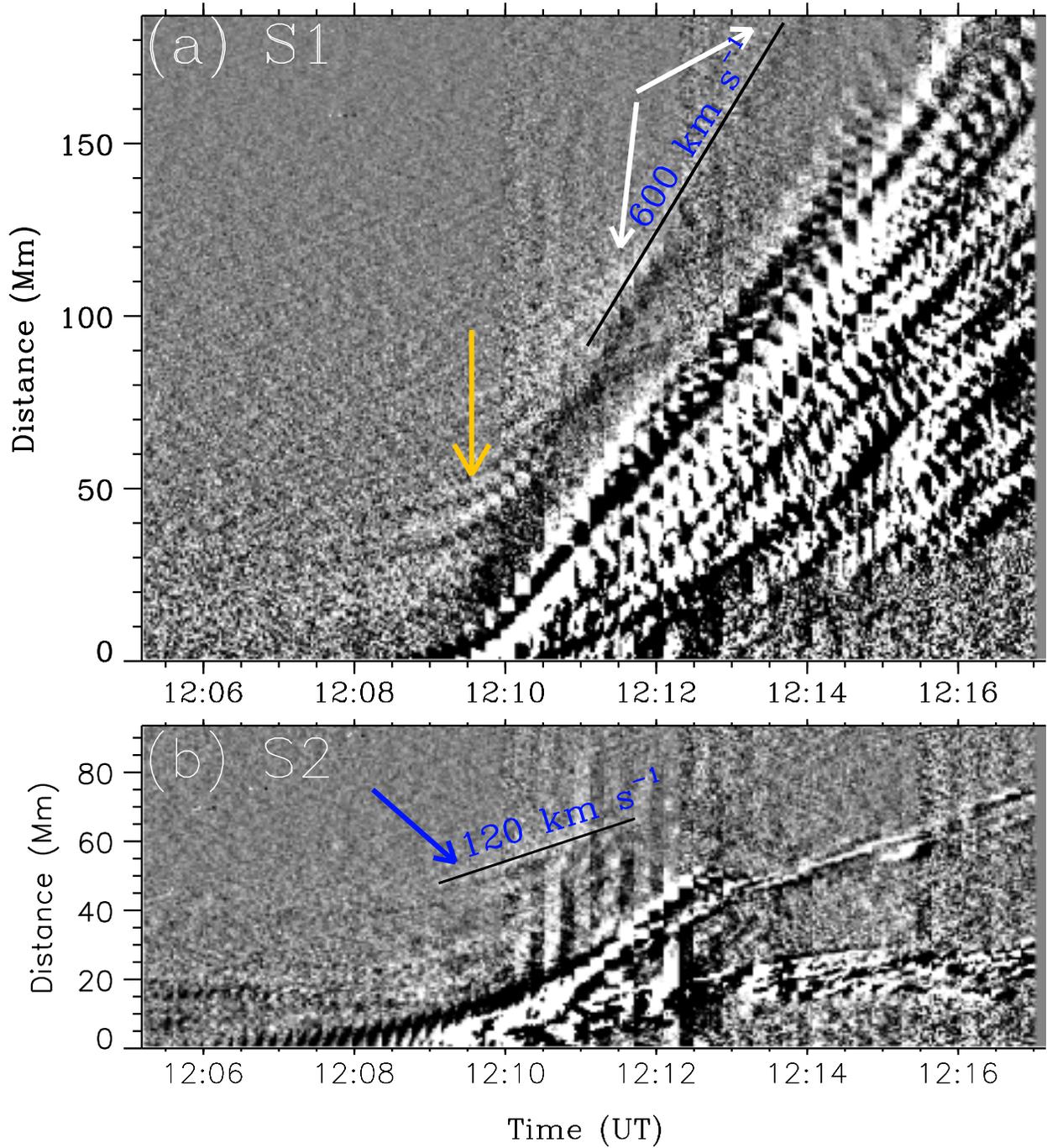}
\caption{Distance-time maps along S1 (a) and S2 (b). The two
arrows in (a) point to the multi-\textbf{shell} (yellow) and
single-\textbf{shell} (white) structures. The speeds are given by
linear fits of the distance measurements.} \label{Fig:fig2}
\end{figure}

\begin{figure}
 \includegraphics[width=1.0\textwidth]{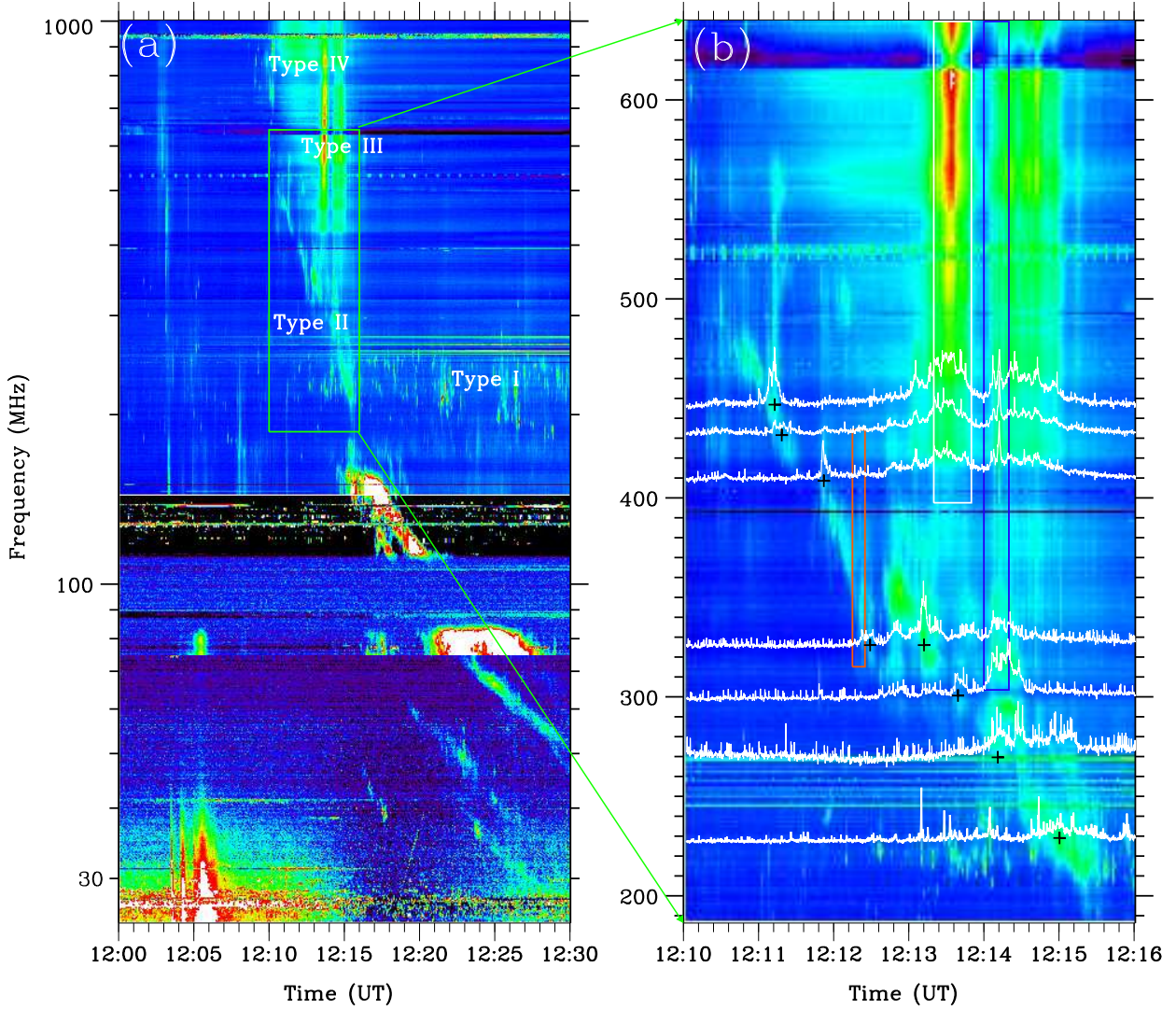}
\caption{Dynamic spectra from ORFEES and SVTO/RSTN (a), with the
box region shown in (b). The 8 pluses on the type II backbone in
(b) indicate the normalized NRH flux peaks at different
frequencies, with NRH sources already presented in Figure 1. The
bottom level of these flux curves is placed at the corresponding
NRH frequency on the y-axis. The red box around 12:12:20 UT is for
the type II herringbone structure (see Figure 4), and the white
and blue boxes around 12:13:30 and 12:14:10 UT are for the
reversely drifting type III bursts (see Figure 5).}
\label{Fig:fig3}
\end{figure}

\begin{figure}
 \includegraphics[width=1.0\textwidth]{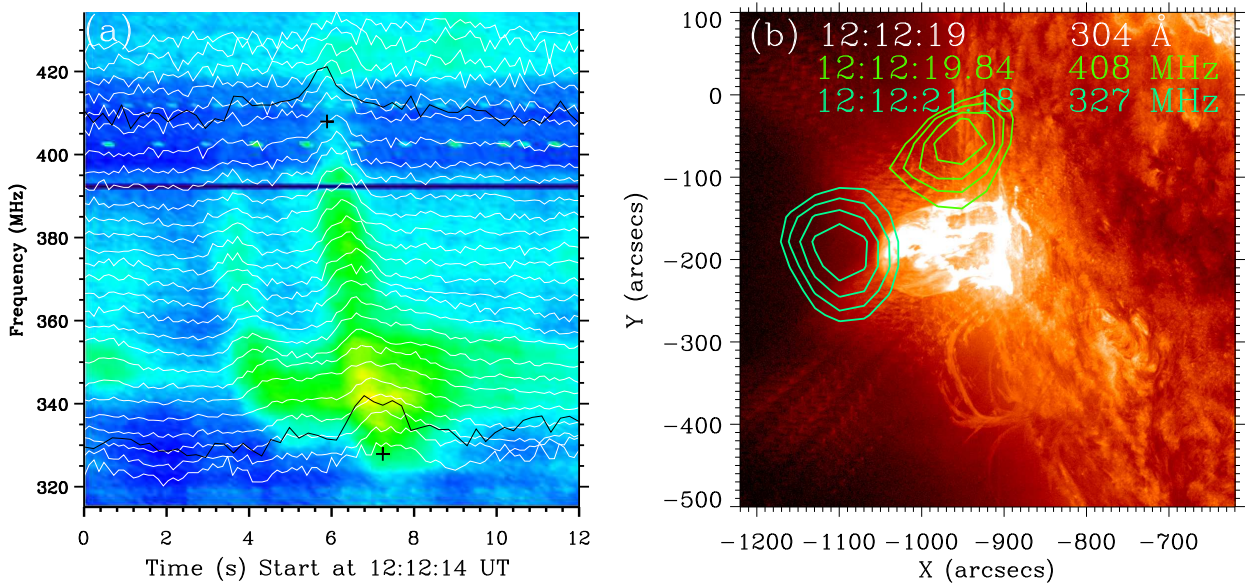}
\caption{The spectral and imaging data for the type II herringbone
(see the red box in Figure 3b). (a) The dynamic spectrum,
over-plotted by the normalized $T_B$ curves for the spectral data
(white) and NRH fluxes at 408 and 327 MHz (black). The bottom
levels of the curves are placed at the corresponding NRH frequency
on the y-axis. The two pluses show the spectral locations of the
radio sources presented in (b) with the AIA 304 {\AA} image.}
\label{Fig:fig4}
\end{figure}

\begin{figure}
 \includegraphics[width=1.0\textwidth]{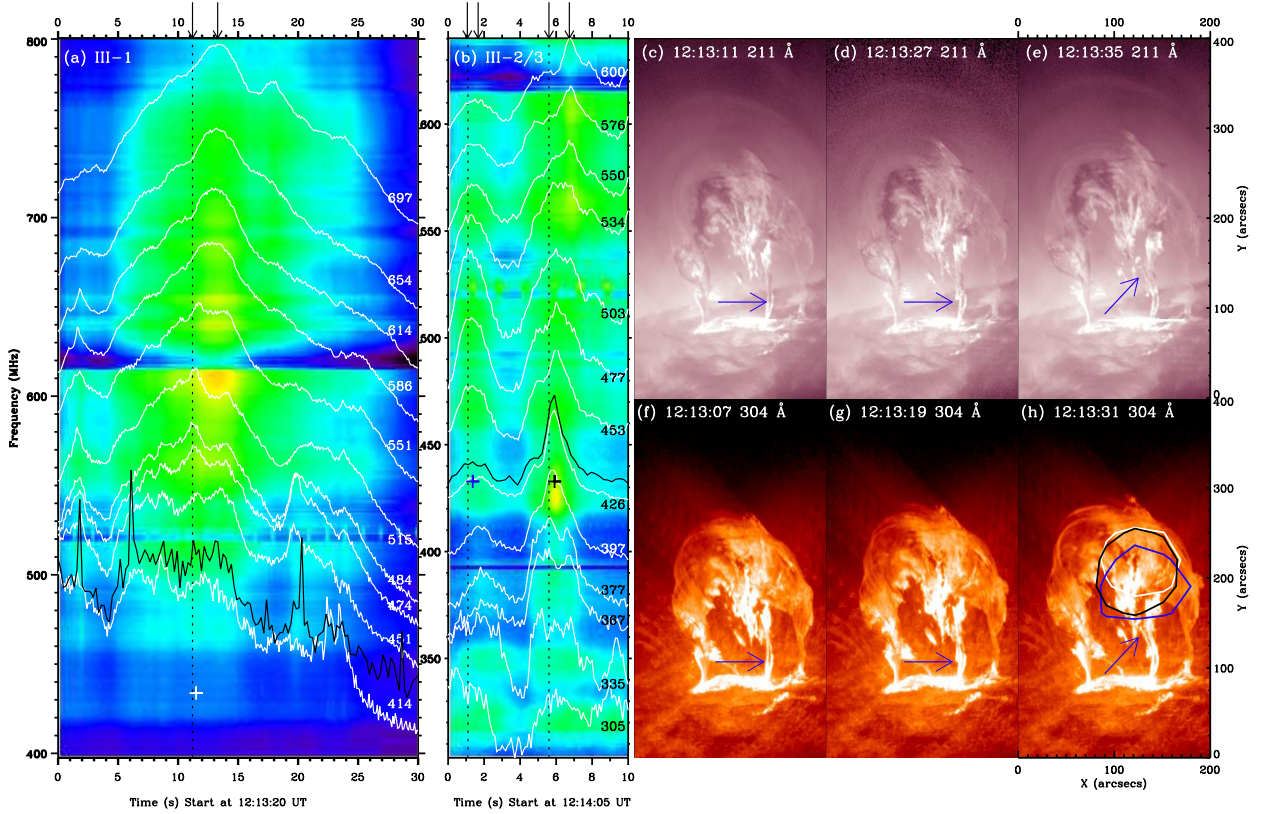}
\caption{ Radio spectral (a-b) and imaging data (c-h) for the
three episodes of type IIIs. (a-b) Their dynamic spectra (see the
white and blue boxes in Figure 3b), over-plotted by the normalized
flux profiles (white) and NRH $T_B$ curves (black). The three
pluses show the selected points at 432 MHz at which the NRH
sources are presented in panel (h). (c-h) The imaging sequence of
AIA at 211 and 304 \AA{} around 12:13 UT, to show the signature of
the suspected reconnection within the ejecta. The arrows point to
the pair of parallel filamentary structures and their morphology
change. The pairs of vertical short arrows present the estimated
durations of the reversely-drifting part of type IIIs. An
accompanying movie is available online.} \label{Fig:fig5}
\end{figure}

\end{document}